\documentclass[pre,aps,twocolumn,showpacs]{revtex4}
\usepackage{graphicx,epsfig,amsmath,amsfonts}

\def\vec#1{{\bf #1}}
\def\text#1{{\mathrm #1}}
\begin{document}

\title{The Lyapunov spectrum of the many-dimensional dilute random Lorentz gas}
\author{Astrid S. de Wijn}
\email{A.S.deWijn@phys.uu.nl}
\author{Henk van Beijeren}
\email{H.vanBeijeren@phys.uu.nl}
\affiliation{Institute for Theoretical Physics, Utrecht University, Leuvenlaan 4, 3584 CE, Utrecht, The Netherlands}

\begin{abstract}
\noindent  For a better understanding of the chaotic behavior of systems of many moving particles it is useful to look at other systems with
many degrees of freedom. An interesting example is the high-dimensional Lorentz gas, which, just like a system of moving hard spheres, may be interpreted as a dynamical system consisting of a point particle in a high-dimensional phase space, moving among fixed scatterers.
In this paper, we calculate the full spectrum of Lyapunov exponents for the dilute random Lorentz gas in an arbitrary number of dimensions.
We find that the spectrum becomes flatter with increasing dimensionality.
Furthermore, for fixed collision frequency the separation between the largest Lyapunov exponent and the second largest one increases logarithmically with dimensionality, whereas the separations between Lyapunov exponents of given indices not involving the largest one, go to fixed limits.
\end{abstract}

\date{\today}

\maketitle

\section{Introduction}
Many studies have been done on the chaotic properties of the Lorentz gas \cite{long1,henkenbob1,henkenbob2}.
It is a simple model which consists of a single particle moving freely between elastic spherical scatterers.
The scatterers can either be placed at random or in a lattice configuration. 
The Lorentz gas provides a physical system yet is still simple enough to allow for exact calculations of many properties.
This simplicity is partially due to the fact that the Lorentz gas contains only one moving particle, and therefore does not have many dynamical degrees of freedom.
Systems with more degrees of freedom, such as moving hard spheres or disks, have also been studied frequently. Extensive simulation work has been done on their Lyapunov spectrum \cite{posch1,forster,christina}, and for low densities analytic calculations have been done for the largest Lyapunov exponent \cite{prlramses,ramses,leiden,jstatph}, the Kolmogorov-Sinai entropy
\cite{prlramses} and for the smallest positive Lyapunov exponents \cite{mareschal,onszelf}.

From the viewpoint of dynamical systems theory the Lorentz gas and hard sphere systems are very similar, as noted already many years ago by Sinai \cite{sinai}.
Both systems may be viewed as "billiards", i.e.\ systems consisting of fixed obstacles in a, mostly high-dimensional, configuration space, among which a point particle moves elastically. In the case of the Lorentz gas these scatterers are (hyper)spheres, in that of the hard-sphere system they are hypercylinders. In either case the convexity of the scatterers makes the system strongly chaotic. In several respects the Lorentz gas is much simpler than the hard sphere system. First of all the scatterers for the Lorentz gas are invariant under rotations in configuration space, which, as we will see, simplifies calculations enormously.
Further, the uniform convexity of the Lorentz gas scatterers, in contrast to the hypercylinders of the hard-sphere systems, strongly simplifies proofs of ergodic and chaotic properties
\cite{szasz-simanyi}. Yet we think it is of interest to perform an explicit calculation of the full Lyapunov spectrum of a high-dimensional dilute Lorentz gas. It is interesting to see the similarities as well as the differences
between the Lorentz gas and the hard-sphere spectra. In addition the methods used here may well be amenable
to refinements, so as to make them applicable to systems of many moving particles.

In this paper, we study the behavior of a dilute, random, non-overlapping, Lorentz gas in an arbitrary number of dimensions, $d$.
For large $d$ this system has many degrees of freedom, while, largely due to the spherical symmetry of the scatterers, it is still possible to do exact calculations. Here we
calculate the full Lyapunov spectrum in the absence of any external fields.

As a preparation we introduce Lyapunov exponents and the Lorentz gas in section \ref{sec:lorentz} and we discuss the low density approximation.
In section \ref{sec:psf} we derive an integral expression for the spectrum of Lyapunov exponents.
Then, in section \ref{sec:spectrum}, we approximate this expression for high-dimensional systems and investigate its properties.
In section \ref{sec:discussion} we discuss the results and make comparisons to the hard-disk Lyapunov spectrum.

\section{\label{sec:lorentz}Lyapunov exponents and the Lorentz gas}
Consider a system with an $\cal N$-dimensional phase space $\Gamma$.
At time $t=0$ the system is at an initial point ${\vec\gamma}_0$ in this space.
It evolves with time, according to ${\vec\gamma}({\vec\gamma}_0,t)$.
If the initial conditions are perturbed infinitesimally, by $\delta{\vec\gamma}_0$, the system evolves along an infinitesimally different path $\gamma + \delta \gamma$, specified by
\begin{eqnarray}
{\delta{\vec\gamma}({\vec\gamma}_0,t)} \label{eq:M}= {{\sf M}_{{\vec\gamma}_0}(t)\cdot \delta{\vec\gamma}_0~,}
\end{eqnarray}
in which the matix $ {\sf M}_{{\vec\gamma}_0}(t)$ is defined by
\begin{eqnarray}
 \label{eq:matrix} \label{eq:tang} {\sf M}_{{\vec\gamma}_0}(t)=\frac{d {\vec\gamma}({\vec\gamma}_0,t)}{d {\vec\gamma}_0}~.
\end{eqnarray}
The Lyapunov exponents are the possible average rates of growth of such perturbations, i.e.,
\begin{equation}
\lambda_i = \lim_{t\rightarrow\infty} \frac{1}{t} \log 
|\mu_i(t)|~,
\end{equation}
where $\mu_i(t)$ is the $i$-th eigenvalue of ${\sf M}_{{\vec\gamma}_0}(t)$.
If the system is ergodic, it comes arbitrarily close to any point in phase space for all initial conditions except for a set of measure zero.
Therefore, the Lyapunov exponents are the same for almost all initial conditions.
We will order the exponents according to size, with $\lambda_1$ being the largest and $\lambda_{\cal N}$ the smallest, as is the convention.
For each exponent there is a corresponding eigenvector of ${\sf M}_{{\vec\gamma}_0}(t)$.


The dynamics of a purely Hamiltonian system
are completely invariant under time reversal.
Also, for ergodic Hamiltonian systems, due to the incompressibility of flow in phase space, the phase-space attractor is invariant under time reversal.
Therefore, every tangent-space eigenvector that grows exponentially under time evolution, shrinks exponentially under backward time evolution. As a consequence,
since the Lyapunov spectrum does not change under time reversal, there is a negative exponent of equal absolute value for every positive Lyapunov exponent.
This is called the conjugate pairing rule.

\subsection{The Lorentz gas}
We will consider the dilute random Lorentz gas.
It is a system with a fixed number $N$ of randomly placed spherical scatterers with diameter $a$ at a (small) density $n$ in $d$ dimensions.
To be specific, we will restrict ourselves to the case of non-overlapping scatterers, where, a priori, each configuration
without overlap among any two scatterers is equally likely.
We will assume that the system is large.
As long as it is finite our preferred boundary conditions are periodic ones, but our considerations allow taking the infinite system limit, at fixed $n$, without any problem.
A single point particle moves between the scatterers, with speed $v$, undergoing a specular reflection at each collision.
For given scatterer positions, the phase space is represented by the position and velocity of the point particle, $\gamma = (\vec{r}, \vec{v})$.
The tangent phase space at any point in phase space can be represented by the perturbations in these quantities, $\delta\gamma = (\delta\vec{r}, \delta\vec{v})$.

\begin{figure}
\includegraphics[width=6cm]{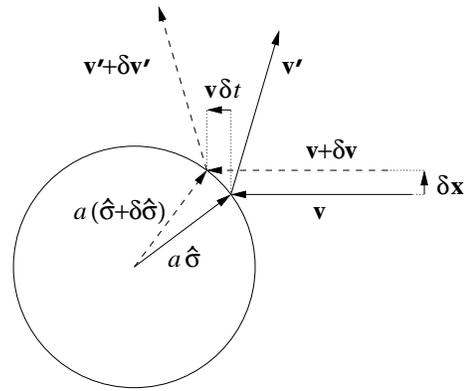}
\caption{Geometry of a collision. The collision normal $\hat{\vec{\sigma}}$ is the unit vector pointing from the center of 
the scatterer to the moving point particle.
\label{fig:lorentzdiagrammetje}}
\end{figure}

Between collisions with the scatterers, the particle moves freely, so $\dot{\vec{r}} = \vec{v}$ and $\dot{\vec{v}} = 0$.
At a collision with collision normal $\hat\sigma$, the particle is reflected by a scatterer, as shown in Fig.~\ref{fig:lorentzdiagrammetje}, with resulting velocity $\vec{v}' = ({
\bf 1} -2 \hat\sigma \hat\sigma) \cdot \vec{v}$.
Here ${
\bf 1}$ is the $d\times d$ identity matrix.
From Eq.~(\ref{eq:matrix}) the temporal behavior of $\delta\gamma$ can be derived.
During free flight,
\begin{eqnarray}
\delta\dot {\vec{r}} &=& \delta\vec{v}~,\\
\delta\dot{\vec{v}} &=& 0~.
\end{eqnarray}
At a collision,
$\delta\vec{v}$ is also reflected.
On the tangent trajectory, the perturbation of the precollisional position with respect to the reference trajectory, leads to a perturbation in the collision normal and the collision time, which shows up in the perturbation of the postcollisional velocity. The perturbation of the collision time also leads to a slight
deviation of the postcollisional position difference (measured at the instant of collision on the tangent trajectory) from the precollisional one (measured at the instant of collision on the reference trajectory). As a result, the 
tangent phase space vectors are transformed according to \cite{ramses,soft2}
\begin{eqnarray}
\label{eq:vlucht}\delta\vec{r}' &=& ({
\bf 1} -2 \hat\sigma \hat\sigma) \cdot \delta\vec{r}~,\\
\delta\vec{v}' &=& ({
\bf 1} -2 \hat\sigma \hat\sigma) \cdot \delta\vec{v} 
\label{eq:bots}
+ 2 {\sf Q} \cdot \delta\vec{r}~,
\end{eqnarray}
in which the collision matrix ${\sf Q}$ is defined by
\begin{eqnarray}
\label{eq:Q}
{\sf Q} =  \frac{1}{a} \left(\vec{v} \hat\sigma - \hat\sigma \vec{v} + \frac{\vec{v}^2}{\vec{v}\cdot\hat\sigma} \hat\sigma \hat\sigma - (\vec{v} \cdot \hat\sigma ) {
\bf 1}\right) ~.
\end{eqnarray}
From these equations it follows that, if $\delta\vec{r}$ and $\delta\vec{v}$ are both parallel to $\vec{v}$ before the collision, $\delta\vec{r}'$ and $\delta\vec{v}'$ are parallel to $\vec{v}'$ after the collision and their absolute values are the same as before.
These are two linearly independent perturbations, giving rise to two zero Lyapunov exponents. They result from time translation invariance and from invariance of the trajectories in configuration space under a scaling of the velocity. All other Lyapunov exponents for the Lorentz gas are non-zero. As a consequence of the conjugate pairing rule $d-1$ of them are positive and the remaining $d-1$ are negative with the same absolute values.

\subsection{The largest Lyapunov exponent at low densities\label{subsec:lown}}
To calculate the non-zero Lyapunov exponents, one needs to consider what happens to an initial perturbation $\delta
\gamma(0)$ in tangent space in the limit of infinite time.
As an introduction we first review the calculation of the largest Lyapunov exponent at low scatterer densities.

An initial perturbation which is not parallel to $\vec{v}$ generically has a non-vanishing component along the most rapidly growing eigenvector of the time evolution operator in tangent space.
Therefore its evolution for long times will be dominated by the largest Lyapunov exponent.
To calculate this time evolution it suffices to consider the growth of the projection of the growing vector onto a subspace of tangent space.
It turns out convenient using the projection onto $\delta \vec{v}$ for this.

Define $\delta\vec{r}_i$ and $\delta\vec{v}_i$ as the tangent space vectors just after collision $i$, with collision normal $\hat\sigma_i$, occurring a time $\tau_i$ after collision $i-1$.
Let $\theta_i$ be defined by 
\begin{equation}
\hat\sigma_i \cdot \vec{v}= 
-v \cos\theta_i.
\end{equation}
Before collision $i$ one has 
\begin{equation}
\delta\vec{r}_i^- = \delta{\vec{r}}_{i-1} + \tau_i \delta\vec{v}_{i-1},
\label{deltar}
\end{equation}
where $\delta\vec{r}_i^-$ is used to indicate the perturbation in the position just before collision $i$.
From Eqs.~(\ref{eq:bots}) and (\ref{eq:Q}) it follows that, typically, after the $(i-1)$-th collision $\delta\vec{v}_{i-1}$ is of the order of $\delta\vec{r}_{i-1} v/a$.
At low densities the mean free time is
of the order of $1/(n a v)$. Therefore, to leading order in the density the first term on the right hand side of Eq.~(\ref{deltar}) may be neglected.
Similarly, in Eq.~(\ref{eq:bots}) the first term on the right hand side becomes negligible at low density, and Eqs.~(\ref{deltar}) and (\ref{eq:bots})
may be combined into
\begin{eqnarray}
\label{eq:benadering}\delta\vec{v}_i = 2\tau_i {\sf Q}_i \cdot \delta\vec{v}_{i-1}~.
\end{eqnarray}
The contributions to the Lyapunov exponents of the terms neglected in this approximation are at least one order of $n$ higher than the terms of leading order \cite{long1}.
In addition, in this approximation, the time reversal symmetry has been destroyed,
hence only the positive Lyapunov exponents can be calculated.
However, for the limit of density going to zero, the results for these will be exact.

The action of ${\sf Q}_i$ on $\delta\vec{v}_i$ may be described in the following way. Working on
the component along $\vec{v}_i$ it yields zero.
It multiplies the component normal to $\vec{v}$ in the plane through $\vec{v}_i$ and $\hat\sigma$ with a factor $v/(a \cos\theta_i)$ and rotates it to the direction in this plane normal to ${\vec{v}'}_i$. Finally it multiplies all other components of 
$\delta\vec{v}_i$ with $v\cos\theta_i/a$.
Define the unit vector orthogonal to $\vec{v}$ in the plane spanned by $\vec{v}$ and $\hat\sigma$ as
\begin{eqnarray}
\hat \rho_i&=&\frac{({\bf 1}-\hat v_i \hat v_i)\cdot \hat\sigma_i}{|\sin\theta_i|}.
\label{eq:rho_i}
\end{eqnarray}
One may rewrite ${\sf Q}_i$ as
\begin{eqnarray}
{\sf Q}_i&=&\frac v a \left(\cos\theta_i({\bf 1}-\hat v_i \hat v_i-\hat\rho_i\hat\rho_i)+\frac 1 {\cos\theta_i} \hat{\rho}'_i\hat{\rho}_i\right).
\end{eqnarray}
Combining this with Eqs.~(\ref{eq:vlucht},\ref{eq:bots}) one finds that the velocity deviations to leading order in $n$ evolve according to 
\begin{eqnarray}
\label{eq://v}\delta\vec{v}_i = \frac{2 v \tau_i}{a} \bigg[\cos\theta_i({
\bf 1} - \hat{\vec{v}}_i\hat{\vec{v}}_i - \hat{\rho}_i \hat{\rho}_i )\nonumber\\
+ \frac{1}{\cos\theta_i} \hat{\rho}'_i \hat{\rho}_i \bigg] \cdot \delta\vec{v}_{i-1}~.
\label{eq:TTvsigma}
\end{eqnarray}
The largest Lyapunov exponent may now be calculated to leading order in the density, as
\begin{eqnarray}
\lambda_1=\bar\nu_d\left\langle\log\frac{|\delta\vec{v}_i|}{|\delta\vec{v}_{i-1}|}\right\rangle~,
\end{eqnarray}
where $\bar\nu_d$ is the average collision frequency for the system and the brackets indicate an average over the collision sequence, which will be discussed in more detail in Sec.~\ref{sec:psf}.
In 2 and 3 dimensions these calculations have been done before~\cite{long1}. The result
in $d$ dimensions will appear as a special case of the calculations presented in the next section.

\section{\label{sec:psf}Partial stretching factors}
In standard terminology, the stretching factor is defined as the factor by which the expanding part of tangent space expands over a time $t$.
This quantity can be used to calculate the Ruelle pressure as well as the sum of the positive Lyapunov exponents, equaling the Kolmogorov-Sinai entropy in systems without escape \cite{henkenbob1,henkenbob2}.

We may define the {\em partial stretching factor} $\Lambda_S(\vec{r},\vec{v}, t)$ of a $p$-dimensional subspace $S$ of the $2d$-dimensional tangent phase space as the factor by which the volume of an infinitesimal $p$-dimensional hypercube in this subspace has increased after a time $t$.
Unless $S$ is orthogonal to some eigenvector associated with one of the $p$ largest Lyapunov exponents, the partial stretching factor for very long times will be dominated by the $p$ most unstable directions in tangent phase space, in other words, by the $p$ largest Lyapunov exponents.
Explicitly, one has the identity
\begin{equation}
\sum_{i=1}^p \lambda_i = \lim_{t\rightarrow \infty} \frac{1}{t} \log\Lambda_{S}(\vec{r},\vec{v}, t)~.\label{eq:psf}
\end{equation}
As in the case of the largest Lyapunov exponent, where we could consider the long time growth of a
basically arbitrary vector in tangent space, we may choose the subspace $S$ in the way that is most convenient to us.
And, as before, we choose $S$ as a subspace of the space spanned by velocity deviations perpendicular to $\vec{v}$.

The partial stretching factor just after collision $N$ is the product of the partial stretching factors of the collisions $1$ through $N$.
These depend on the relative orientations of $\vec{v}$, $\hat\sigma$, and the image of $S$.
One can write
\begin{eqnarray}
\Lambda_S(\vec{r},\vec{v}, t_N)&=&\prod_{i=1}^N \Lambda_p^{(i)}(v, \tau_i, \theta_i,\alpha_i)~.\label{eq:product}~,
\end{eqnarray}
in which $\alpha_i$ is the projection angle of $\hat\rho_i$ onto the image of $S$ after the $(i-1)$-th collision.
The subspace $S$ can be split into a $(p-1)$-dimensional subspace normal to $\hat\rho_i$ and a $1$-dimensional subspace spanned by the projection of $\hat\rho_i$ onto $S$.
From Eq.~(\ref{eq:TTvsigma}) one finds that the former contributes a factor of $(\tau_i\cos\theta_i)^{p-1}$ to the partial stretching factor.
The projection of $\hat\rho_i$ onto the image of $S$ can be split into components perpendicular and parallel to $\hat\rho_i$.
The former grows with $2 v \tau_i\cos\theta_i/a$ and the latter with $2 v \tau_i/(a \cos\theta_i)$.
The partial stretching factor thus becomes
\begin{eqnarray}
\lefteqn{\Lambda_p^{(i)}(v, \tau_i, \theta_i,\alpha_i) = }&&\nonumber\\
 && \left({\displaystyle \frac{2 v \tau_i}{a}}\right)^{p} \cos^{p-1}\theta_i \nonumber\\
&\phantom{=}&\times\sqrt{(\sin\alpha_i\cos\theta_i)^2 + \left({\displaystyle \frac{\cos\alpha_i}{\cos\theta_i}}\right)^2}~.
\label{eq:factor}
\end{eqnarray}
From this expression one can calculate the Lyapunov exponents and obtain asymptotic approximations for high dimensionality.

\subsection{Lyapunov exponents}
The sum of the $p$ largest Lyapunov exponents can be calculated
by substituting Eq.~(\ref{eq:product}) into Eq.~(\ref{eq:psf}),
\begin{eqnarray}
\sum_{i=1}^{p} \lambda_i 
&=& \lim_{t_N\rightarrow \infty} \frac{1}{t_N} \sum_{i=1}^N \log \Lambda_p^{(i)}(v, \tau_i, \theta_i,\alpha_i)~,\label{eq:somlog}
\end{eqnarray}
where $t_N$ is the time at which the $N$-th collision occurs.

In the low density limit, the collisions in the Lorentz gas are uncorrelated and therefore the time average in Eq.~(\ref{eq:somlog}) can be replaced with an ensemble average,
\begin{eqnarray}
\label{eq:ergodic}\sum_{i=1}^{p} \lambda_i&=& \int_0^\infty\! d\tau \int_0^{\frac{\pi}{2}}\!d\theta \int_0^{\frac{\pi}{2}}\! d\alpha\nonumber\\
&&\null \times \nu_p(\tau,\theta,\alpha;v) \log \Lambda_p^{(i)}(v, \tau, \theta,\alpha)~.
\end{eqnarray}
Here $\nu_p(\tau,\theta,\alpha;v)$ is a probability distribution, describing the probability density per units of time and angle for collisions with the parameters $\tau, \theta$, and $\alpha$ at a given velocity $v$.
The index $p$ is attached to remind of the dependence of the distribution of $\alpha$ on the dimensionality of the subset $S$.
A more rigorous proof of the theorem used to derive Eq.~(\ref{eq:ergodic}) can be found in Crisanti, Paladin, and Vulpiani \cite{drieitalianen}.

\subsection{The distribution of free flight times}
To lowest order in the density of scatterers collisions are uncorrelated; effects of recollisions
appear in the Lyapunov exponents only at higher orders.
With increasing $d$ these higher density corrections even become smaller, since the probability of a return to a scatterer decreases rapidly.
In this approximation the time of free flight between consecutive collisions is distributed exponentially. In addition the distribution of the angles $\theta$ and $\alpha$ is independent of that of the free flight time, the direction of the incident velocity and the orientation of the precollisional image of $S$. 
The probability density for colliding at angle $\theta$ is proportional to the differential cross section. In the free flight time distribution it contributes a factor of
\begin{eqnarray}
\label{eq:vol} n a^{d-1} v\, O_{d-1} \sin^{d-2}\theta\cos\theta\, d\theta dt~.
\end{eqnarray}
Here $O_m$ is the $(m-1)$-dimensional surface area of the $m$-dimensional unit sphere, i. e.,
\begin{eqnarray}
O_m &=& \frac{2\, \pi^{\frac{m}{2}}}{\Gamma\left(\frac{m}{2}\right)}~.
\end{eqnarray}
Finally, the probability distribution of the projection angle $\alpha$ of $\hat{\rho}$ onto the image of $S$ after $i-1$ collisions may be identified with the fraction of the $(d-1)$-dimensional unit sphere that has a projection angle between $\alpha$ and $\alpha + d\alpha$, viz.\
\begin{eqnarray}
\label{eq:rho} \rho(\alpha)= \begin{cases}
{\displaystyle \frac{O_{d-1-p} O_p}{O_{d-1} }} \sin^{d-2-p}\alpha  \cos^{p-1}\alpha&{\rm if~} p<d-1\\
\delta(\alpha) &{\rm if~} p=d-1
\end{cases}
~,
\end{eqnarray}
where $\delta(\alpha)$ is the Dirac delta function.
Combining Eqs.~(\ref{eq:vol}) and (\ref{eq:rho}) with the exponential distribution of the free flight times, one finds, for $p<d-1$,
\begin{eqnarray}
\lefteqn{\nu_p(\tau,\theta,\alpha;v)= }&&\nonumber\\
 && 
 2n a^{d-1} v \,  {O_{d-1-p}\, O_p} \,\bar\nu_d\, \exp-{\bar\nu_d\tau}\nonumber\\
&&\null\times\label{eq:nu} \sin^{d-2}\theta\,\cos\theta\, \sin^{d-2-p}\alpha\, \cos^{p-1}\alpha ~,
\end{eqnarray}
where the average collision frequency has the explicit form
\begin{eqnarray}
\bar\nu_d &=& \frac{n a^{d-1} v\, O_{d-1}}{d-1}~.
\end{eqnarray}
Note that $O_{d-1}/(d-1)$ is equal to the volume of the $(d-1)$-dimensional unit
sphere, so $a^{d-1}O_{d-1}/(d-1)$ is the total cross-section for a collision with a scatterer.

\section{The spectrum\label{sec:spectrum}}
Substituting Eqs.~(\ref{eq:factor}) and (\ref{eq:nu}) into Eq.~(\ref{eq:ergodic}) yields an expression for the sum of the $p$ largest Lyapunov exponents,
\begin{widetext}
\begin{eqnarray}
\sum_{i=1}^{p} \lambda_i&=&
\int_0^\infty\! d\tau \int_0^{\frac{\pi}{2}}\! d\theta\, n a^{d-1} v \,  {O_{d-1}} \bar\nu_d \exp(-{\bar\nu_d\tau})\sin^{d-2}\theta\cos\theta
\left[ p \log\! \left(\frac{2 v \tau}{a}\right)+ (p-1) \log \cos\theta \right]\nonumber\\
&&\null+{\textstyle
}
\int_0^{\frac{\pi}{2}}\! d\theta \int_0^{\frac{\pi}{2}}\!d\alpha\,
n a^{d-1} v \,  {O_{d-1-p} O_p} 
\sin^{d-2}\theta\cos\theta \sin^{d-2-p}\alpha \cos^{p-1}\alpha
\log\!\left[(\sin\alpha\cos\theta)^2 + \left(\frac{\cos\alpha}{\cos\theta}\right)^2\right]~.
\label{eq:int}
\end{eqnarray}
\end{widetext}
The first integral can easily be carried out analytically.  The second integral is more difficult.
It can be simplified in special cases, as described in subsection \ref{subsec:ks}, or approximated, as described in the remainder of this section.

\subsection{The KS entropy\label{subsec:ks}}
By taking $p=d-1$, one can calculate the sum of all the positive exponents, the Kolmogorov-Sinai (KS) entropy.
Since the distribution of $\alpha$ now is a delta function, the second term in Eq.~\ref{eq:int} becomes
\begin{eqnarray}
-\int_0^{\frac{\pi}{2}}\! d\theta \,
n a^{d-1} v \,  {O_{d-1-p} O_p} 
\sin^{d-2}\theta\cos\theta \nonumber\\
\sin^{d-2-p}\alpha\, \cos^{p-1}\alpha\,
\log\,\cos\theta
\end{eqnarray}

Performing the integrals yields,
\begin{eqnarray}
\sum_{i=1}^{d-1} \lambda_i &=& \frac
{a\bar{\nu}_d}{2v}
\nonumber\\
&&\times \Bigg\{
2 (d-1) \left[ - \log\! \left(\frac{n a^d O_{d-1}}{2(d-1)}\right) - \gamma \right]
\nonumber\\
&&\null - (d-3) \left[\psi^{(0)}\!\left(\frac{d+1}{2}\right)+\gamma\right]
\Bigg\}~.\label{eq:ks}
\end{eqnarray}
Here $\psi^{(i)}(x)$ is the $(i+1)$-th derivative of the $\log\Gamma(x)$; $\psi^{(i)} (x) = [\log \Gamma(x)]^{(i+1)}$.
This reproduces the results of Van Beij\-eren, Latz, and Dorfman \cite{long1,henkenbob1,henkenbob2}.

\subsection{Lower bound\label{subsec:lowerbound}}
A lower bound for the Lyapunov exponents can be derived by assuming that the minimum growth of the tangent space vector is realized for every collision.
The minimum growth is the growth in the directions perpendicular to $\hat\sigma$ and $\vec{v}$.
This yields,
\begin{eqnarray}
\lambda_{d-1}&>& \lambda_-~,
\end{eqnarray}
with $\lambda_-$  defined by
\begin{eqnarray}
\lambda_- &=& \frac{\bar\nu_d}{2}
\bigg[ - 2 \log \!\left(\frac{
a\bar\nu_d}{2
v}\right)\nonumber\\
&&\phantom{\frac{\bar\nu_d}{2}[} \null- 3 \gamma  - \psi^{(0)}\!\left(\frac{d+1}{2}\right)\bigg] ~.
\end{eqnarray}
The function $\psi^{(0)}({d+1}/{2})$ for large $d$ behaves as $\log d$.
Therefore the $\log \bar\nu_d$ term dominates the behavior of $\lambda_-$, and
$\lambda_-$ for large $d$ behaves asymptotically as $\bar\nu_d \log \bar\nu_d$.

In a system of high dimensionality, the exponents are dominated by the directions perpendicular to $\vec{v}$ and $\hat\sigma$.
They should behave as $\lambda_-$ with a small correction.

\subsection{High dimensionality\label{subsec:highd}}
The integrals in Eq.~(\ref{eq:int}) can be estimated for large $d$ and arbitrary $p$.
For large numbers of dimensions the distribution of $\theta$ is sharply peaked near $\theta = \pi/2$.
The argument of the logarithm in 
the second term of Eq.~(\ref{eq:int}) is therefore dominated by the $1/\cos^2\theta$ term.
The range of $\alpha$ where the other term dominates is of the order of $1/d$, and therefore that term can be neglected.
This leads to
\begin{widetext}
\begin{eqnarray}
\sum_{i=1}^{p} \lambda_i&\approx& p \lambda_- +
\int_0^{\frac{\pi}{2}}d\theta \int_0^{\frac{\pi}{2}}d\alpha\,
n a^{d-1} v \,  {O_{d-1-p} O_p}
\sin^{d-2}\theta\,\cos\theta \,\sin^{d-2-p}\alpha\, \cos^{p-1}\alpha\,
\left(- 2 \log \cos\theta + \log\cos\alpha \right),
\end{eqnarray}
which yields, after performing the integrals,
\begin{eqnarray}
\sum_{i=1}^{p} \lambda_i&\approx& p \lambda_- + \frac{\bar\nu_d}{2} \left[2\gamma - \psi^{(0)}\!\left(\frac{d-1}{2}\right) + 2 \psi^{(0)}\!\left(\frac{d+1 }{2}\right) + \psi^{(0)}\!\left(\frac{p}{2}\right)\right]~.
\end{eqnarray}
\end{widetext}

For the largest exponent this means that the behavior for large $d$ is approximately
\begin{eqnarray}
\label{eq:lambda1}\lambda_1 \approx \lambda_- +{\textstyle\frac{1}{2}}  \bar\nu_d \left(\log d +\gamma - {\textstyle \frac{5}{2}} \log 2 -{\textstyle \frac{1}{2}}\right) ~.
\end{eqnarray}
This behavior is shown in the uppermost curve in Fig.~\ref{fig:alles}.

The smallest exponent is equal to the KS-entropy minus the sum of the first $d-2$ exponents.
In the high dimensionality limit the smallest positive exponent behaves as,
\begin{eqnarray}
\label{eq:lambdad-1}\lambda_{d-1} \approx \lambda_- + \frac{\bar\nu_d}{2 d}~.
\end{eqnarray}
This is illustrated in Fig.~\ref{fig:kleinste}
The $p$-th exponent can be calculated by subtracting the expression for the sum of the first $p-1$ exponents from that for the first $p$ ones. This results in
\begin{eqnarray}
\label{eq:lambdap}\lambda_p \approx \lambda_- + \frac{\bar\nu_d}{2} \left[ \psi^{(0)}\!\left(\frac{p}{2}\right) -  \psi^{(0)}\!\left(\frac{p-1}{2}\right) \right]~.
\end{eqnarray}
For the first few exponents this yields
\begin{eqnarray}
\label{eq:lambda2} \lambda_2 &\approx& \lambda_- + \bar\nu_d \log 2~,\\
\label{eq:lambda3} \lambda_3 &\approx& \lambda_- + \bar\nu_d (1-\log 2)~,\\
\label{eq:lambda4} \lambda_4 &\approx& \lambda_- + \bar\nu_d \left(\log 2 - {\textstyle \frac{1}{2}}\right)~,\\
&&\nonumber \dots~.
\end{eqnarray}

As a consequence of this, for fixed collision frequency, $(\lambda_1-\lambda_-)/\bar\nu_d$ grows logarithmically with dimensionality, whereas $(\lambda_p-\lambda_-)/\bar\nu_d$ with $p>1$, approaches a limit that is independent of $d$. This too is illustrated in Fig.~\ref{fig:alles}.

\section{Discussion\label{sec:discussion}}

\begin{figure}
\includegraphics[height=8.6cm,angle=270]{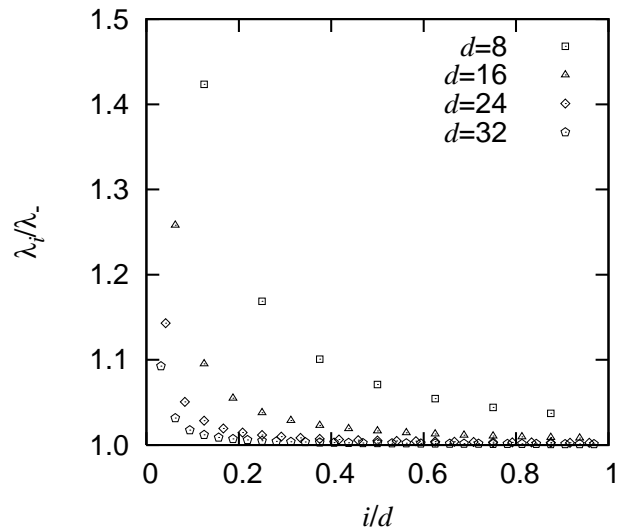}
\caption{The scaled spectrum of positive Lyapunov exponents for several values of the dimensionality $d$ at density $n=0.01 /a^2$.}
\label{fig:spectra}
\end{figure}

\begin{figure}
\includegraphics[height=8.6cm,angle=270]{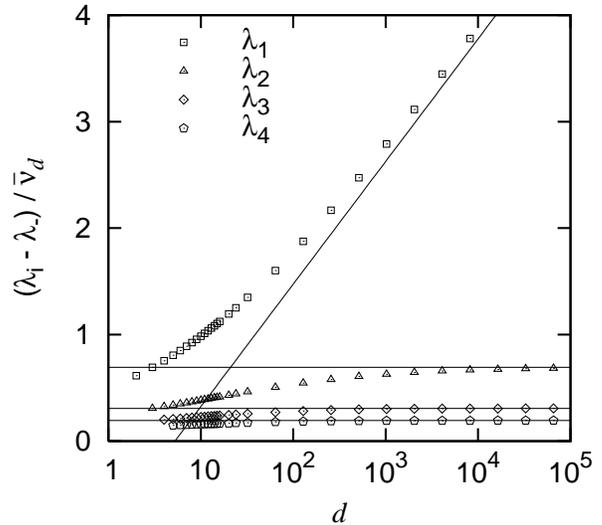}
\caption{The four largest Lyapunov exponents as functions of the dimensionality.  Their limiting behavior is indicated with lines.  The second, third, and fourth exponent converge to horizontal lines, as calculated in Eqs.~(\ref{eq:lambda2}-\ref{eq:lambda4}). }
\label{fig:alles}
\label{fig:grootste}
\label{fig:tweede}
\end{figure}

\begin{figure}
\includegraphics[height=8.6cm,angle=270]{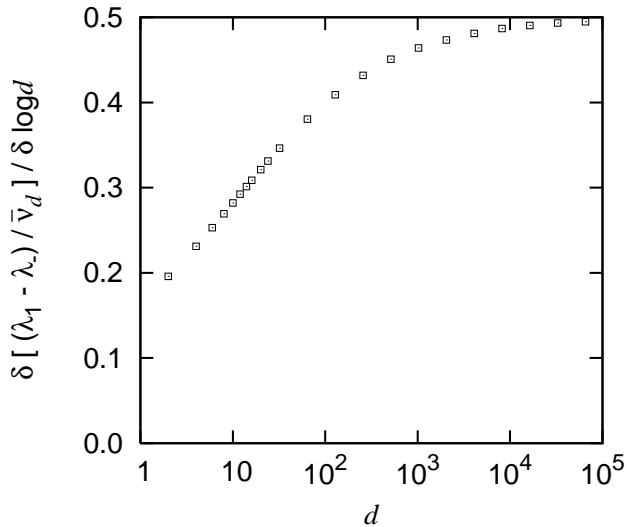}
\caption{The discrete derivative of the largest Lyapunov exponent with respect to $\log d$.  The derivative converges to $\frac{1}{2}$, as calculated in Eq.~(\ref{eq:lambda1}).}
\label{fig:grootstediff}
\end{figure}


\begin{figure}
\includegraphics[height=8.6cm,angle=270]{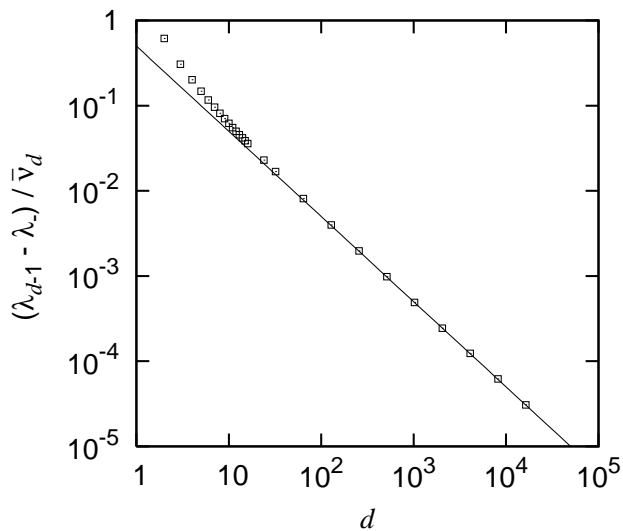}
\caption{The smallest positive Lyapunov exponent as a function of the dimensionality.}
\label{fig:kleinste}
\end{figure}

The integrals in Eq.~(\ref{eq:int}) can be performed numerically.
The results are displayed in Figs.~\ref{fig:spectra}-\ref{fig:kleinste}, along with the limiting behaviors for large dimensionality, which we discussed already in the previous section.

The offset of all Lyapunov exponents,
$\lambda_-$, depends on both density and dimensionality.
For large $d$ its magnitude is determined primarily by dimensionality, unless density is so 
low as to satisfy
\begin{eqnarray}
n a^d  < \frac{O_{d-1}}{d-1}~.
\end{eqnarray}

\begin{figure}
\includegraphics[height=8.6cm,angle=270]{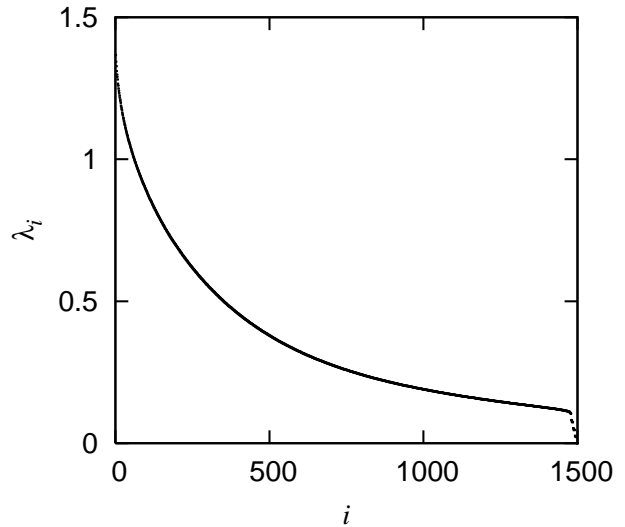}
\caption{\label{fig:harddisks} The spectrum of Lyapunov exponents from simulations \cite{posch1,posch2,christina} of 750 hard spheres in 2 dimensions at density $n=0.1/ a^2$ in a rectangular box of dimensions $10/\sqrt{n} \times 75/\sqrt{n}$, with periodic boundary conditions.
Only the positive exponents are plotted, since, by the conjugate pairing rule, the negative spectrum is exactly the opposite.
The kinetic energy per particle has been scaled to 1.
}
\end{figure}

It is interesting to compare the Lyapunov spectrum of a high dimensional Lorentz gas
to that of a system of moving hard disks, as computed by Dellago et al.~\cite{posch2}. Fig.~\ref{fig:harddisks} shows such a spectrum for 750 particles in 2 dimensions at a density of $n=0.1/a^2$. One immediately notices that the Lorentz gas spectrum is much flatter than that of the hard disk system. The explanation for this is that for the hard disk system only a few (4 to be specific)
components of the velocity in phase space are changed at each collision, whereas in the Lorentz gas all components are involved, though some will be increased more than others. A greater similarity may be obtained by replacing the spherical scatterers in the $d$-dimensional Lorentz gas
by randomly oriented hypercylinders. Indeed our calculations \cite{tbp} indicate that the spectrum obtained for this system resembles much more that of the hard disk system, but there remain significant differences due to the randomness of the cylinder orientations (for hard disk systems the scatterers in phase space are hypercylinders of very specific orientations). In addition, due to the absence of velocities for the scatterers, the $d$-dimensional Lorentz gas will never exhibit a branch of Goldstone modes with Lyapunov exponents approaching zero as the inverse of system size, in contrast to hard disk systems \cite{posch1,mareschal,onszelf}.
However, there are also some similarities between the two (positive) spectra: first of all both become increasingly flat with increasing index (apart from the Goldstone branch for the hard disks).
Secondly both become very steep near the largest exponent.
It is especially remarkable that, for fixed collision frequency, the difference between the largest exponent and the next largest one in the Lorentz gas increases logarithmically with the number of degrees of freedom, whereas all the subsequent differences, in the positive half of the spectrum, approach fixed limiting values.
Whether something like this also happens for hard disks is not known at present, though there have been conjectures of similar behavior by Searles et al~\cite{searles}.

\section{Conclusion}
In this paper we have calculated the full Lyapunov spectrum for the dilute random Lorentz gas in an arbitrary number of dimensions.
We have found analytical expressions for the behavior of the spectrum for systems with many degrees of freedom.
The spectrum becomes flatter with increasing dimensionality or decreasing density.
The separation between the largest and second largest exponent, expressed in units of the collision frequency, increases logarithmically with dimensionality.

Perhaps a similar approach may be applied to the Lyapunov spectra of systems of many particles, such as freely moving hard spheres. However, in the present case we could take advantage of the property that 
the partial stretching factor is distributed independently of the subspace being stretched \cite{drieitalianen}.
This will not be true any more for systems of many moving particles, for which the dynamics are not invariant under arbitrary rotations in configuration space.
Some new ideas will be required here.

\begin{acknowledgments}
We would like to thank
J.~R.~Dorfman for useful discussions,
and C.~Dellago, H.~A.~Posch, R.~Hirschl, and C.~Forster for kindly providing us with their simulation program for hard spheres.
\end{acknowledgments}


\begin{thebibliography}{19}
\expandafter\ifx\csname natexlab\endcsname\relax\def\natexlab#1{#1}\fi
\expandafter\ifx\csname bibnamefont\endcsname\relax
  \def\bibnamefont#1{#1}\fi
\expandafter\ifx\csname bibfnamefont\endcsname\relax
  \def\bibfnamefont#1{#1}\fi
\expandafter\ifx\csname citenamefont\endcsname\relax
  \def\citenamefont#1{#1}\fi
\expandafter\ifx\csname url\endcsname\relax
  \def\url#1{\texttt{#1}}\fi
\expandafter\ifx\csname urlprefix\endcsname\relax\def\urlprefix{URL }\fi
\providecommand{\bibinfo}[2]{#2}
\providecommand{\eprint}[2][]{\url{#2}}

\bibitem[{\citenamefont{{van Beijeren} et~al.}(1998)\citenamefont{{van
  Beijeren}, Latz, and Dorfman}}]{long1}
\bibinfo{author}{\bibfnamefont{H.}~\bibnamefont{{van Beijeren}}},
  \bibinfo{author}{\bibfnamefont{A.}~\bibnamefont{Latz}}, \bibnamefont{and}
  \bibinfo{author}{\bibfnamefont{J.~R.} \bibnamefont{Dorfman}},
  \bibinfo{journal}{Phys.~Rev.~E} \textbf{\bibinfo{volume}{57}},
  \bibinfo{pages}{4077} (\bibinfo{year}{1998}).

\bibitem[{\citenamefont{{van Beijeren} and Dorfman}(1995)}]{henkenbob1}
\bibinfo{author}{\bibfnamefont{H.}~\bibnamefont{{van Beijeren}}}
  \bibnamefont{and} \bibinfo{author}{\bibfnamefont{J.~R.}
  \bibnamefont{Dorfman}}, \bibinfo{journal}{Phys. Rev. Lett.}
  \textbf{\bibinfo{volume}{74}}, \bibinfo{pages}{4412} (\bibinfo{year}{1995}).

\bibitem[{\citenamefont{{van Beijeren} and Dorfman}(2002)}]{henkenbob2}
\bibinfo{author}{\bibfnamefont{H.}~\bibnamefont{{van Beijeren}}}
  \bibnamefont{and} \bibinfo{author}{\bibfnamefont{J.~R.}
  \bibnamefont{Dorfman}}, \bibinfo{journal}{J.~Stat.~Phys.}
  \textbf{\bibinfo{volume}{108}}, \bibinfo{pages}{767} (\bibinfo{year}{2002}),
  \bibinfo{note}{nlin.CD/0112031}.

\bibitem[{\citenamefont{Posch and Hirschl}(2000)}]{posch1}
\bibinfo{author}{\bibfnamefont{H.~A.} \bibnamefont{Posch}} \bibnamefont{and}
  \bibinfo{author}{\bibfnamefont{R.}~\bibnamefont{Hirschl}}, in
  \emph{\bibinfo{booktitle}{Hard Ball Systems and the Lorentz Gas}}, edited by
  \bibinfo{editor}{\bibfnamefont{D.}~\bibnamefont{Szasz}}
  (\bibinfo{publisher}{Springer}, \bibinfo{year}{2000}), Encyclopedia of
  Mathematical Sciences.

\bibitem[{\citenamefont{Forster et~al.}(2004)\citenamefont{Forster, Hirschl,
  Posch, and Hoover}}]{forster}
\bibinfo{author}{\bibfnamefont{C.}~\bibnamefont{Forster}},
  \bibinfo{author}{\bibfnamefont{R.}~\bibnamefont{Hirschl}},
  \bibinfo{author}{\bibfnamefont{H.~A.} \bibnamefont{Posch}}, \bibnamefont{and}
  \bibinfo{author}{\bibfnamefont{W.~G.} \bibnamefont{Hoover}},
  \bibinfo{journal}{Physica~D} \textbf{\bibinfo{volume}{187}}
  (\bibinfo{year}{2004}).

\bibitem[{\citenamefont{Eckmann et~al.}()\citenamefont{Eckmann, Forster, Posch,
  and Zabey}}]{christina}
\bibinfo{author}{\bibfnamefont{J.-P.} \bibnamefont{Eckmann}},
  \bibinfo{author}{\bibfnamefont{C.}~\bibnamefont{Forster}},
  \bibinfo{author}{\bibfnamefont{H.~A.} \bibnamefont{Posch}}, \bibnamefont{and}
  \bibinfo{author}{\bibfnamefont{E.}~\bibnamefont{Zabey}},
  \emph{\bibinfo{title}{Lyapunov modes in hard-disk systems}},
  \bibinfo{note}{nlin.CD/0404007}.

\bibitem[{\citenamefont{{van Zon} et~al.}(1998)\citenamefont{{van Zon}, {van
  Beijeren}, and Dellago}}]{prlramses}
\bibinfo{author}{\bibfnamefont{R.}~\bibnamefont{{van Zon}}},
  \bibinfo{author}{\bibfnamefont{H.}~\bibnamefont{{van Beijeren}}},
  \bibnamefont{and} \bibinfo{author}{\bibfnamefont{C.}~\bibnamefont{Dellago}},
  \bibinfo{journal}{Phys.~Rev.~Lett.} \textbf{\bibinfo{volume}{80}},
  \bibinfo{pages}{2035} (\bibinfo{year}{1998}).

\bibitem[{\citenamefont{{van Zon}}(2000)}]{ramses}
\bibinfo{author}{\bibfnamefont{R.}~\bibnamefont{{van Zon}}}, Ph.D. thesis,
  \bibinfo{school}{Utrecht University} (\bibinfo{year}{2000}).

\bibitem[{\citenamefont{{van Zon} et~al.}(2000)\citenamefont{{van Zon}, {van
  Beijeren}, and Dorfman}}]{leiden}
\bibinfo{author}{\bibfnamefont{R.}~\bibnamefont{{van Zon}}},
  \bibinfo{author}{\bibfnamefont{H.}~\bibnamefont{{van Beijeren}}},
  \bibnamefont{and} \bibinfo{author}{\bibfnamefont{J.~R.}
  \bibnamefont{Dorfman}}, in \emph{\bibinfo{booktitle}{Proceedings of the 1998
  NATO-ASI "Dynamics: Models and Kinetic Methods for Non-equilibrium Many-Body
  Systems}}, edited by
  \bibinfo{editor}{\bibfnamefont{J.}~\bibnamefont{Karkheck}}
  (\bibinfo{publisher}{Kluwer, Dordrecht}, \bibinfo{year}{2000}), pp.
  \bibinfo{pages}{131--167}, \bibinfo{note}{chao-dyn/9906040}.

\bibitem[{\citenamefont{{van Zon} and {van Beijeren}}(2002)}]{jstatph}
\bibinfo{author}{\bibfnamefont{R.}~\bibnamefont{{van Zon}}} \bibnamefont{and}
  \bibinfo{author}{\bibfnamefont{H.}~\bibnamefont{{van Beijeren}}},
  \bibinfo{journal}{J.~Stat.~Phys.} \textbf{\bibinfo{volume}{109}},
  \bibinfo{pages}{641} (\bibinfo{year}{2002}).

\bibitem[{\citenamefont{McNamara and Mareschal}(2001)}]{mareschal}
\bibinfo{author}{\bibfnamefont{S.}~\bibnamefont{McNamara}} \bibnamefont{and}
  \bibinfo{author}{\bibfnamefont{M.}~\bibnamefont{Mareschal}},
  \bibinfo{journal}{Phys.~Rev.~E} \textbf{\bibinfo{volume}{63}}
  (\bibinfo{year}{2001}).

\bibitem[{\citenamefont{{de Wijn} and {van
  Beijeren}}({\natexlab{a}})}]{onszelf}
\bibinfo{author}{\bibfnamefont{A.~S.} \bibnamefont{{de Wijn}}}
  \bibnamefont{and} \bibinfo{author}{\bibnamefont{{van Beijeren}}},
  \emph{\bibinfo{title}{Goldstone modes in lyapunov spectra of hard sphere
  systems}}, \bibinfo{note}{nlin.CD/0312051, accepted for publication in
  Phys.~Rev.~E}.

\bibitem[{\citenamefont{Sinai}(1972)}]{sinai}
\bibinfo{author}{\bibfnamefont{Y.~G.} \bibnamefont{Sinai}},
  \bibinfo{journal}{Russian Mathematical Surveys}
  \textbf{\bibinfo{volume}{27}}, \bibinfo{pages}{21} (\bibinfo{year}{1972}),
  \bibinfo{note}{reprinted in Y.~G.~Sinai, editor, Dynamical Systems, a
  Collection of Papers, World Scientific, Singapore 1991}.

\bibitem[{\citenamefont{Sim\'anyi and Sz\'asz}(1999)}]{szasz-simanyi}
\bibinfo{author}{\bibfnamefont{N.}~\bibnamefont{Sim\'anyi}} \bibnamefont{and}
  \bibinfo{author}{\bibfnamefont{D.}~\bibnamefont{Sz\'asz}},
  \bibinfo{journal}{Annals of Mathematics} \textbf{\bibinfo{volume}{149}},
  \bibinfo{pages}{35} (\bibinfo{year}{1999}).

\bibitem[{\citenamefont{Dellago et~al.}(1996)\citenamefont{Dellago, Posch, and
  Hoover}}]{soft2}
\bibinfo{author}{\bibfnamefont{C.}~\bibnamefont{Dellago}},
  \bibinfo{author}{\bibfnamefont{H.~A.} \bibnamefont{Posch}}, \bibnamefont{and}
  \bibinfo{author}{\bibfnamefont{W.~G.} \bibnamefont{Hoover}},
  \bibinfo{journal}{Phys.~Rev.~E} \textbf{\bibinfo{volume}{53}},
  \bibinfo{pages}{1485} (\bibinfo{year}{1996}).

\bibitem[{\citenamefont{Cristanti et~al.}(1993)\citenamefont{Cristanti,
  Paladin, and Vulpiani}}]{drieitalianen}
\bibinfo{author}{\bibfnamefont{A.}~\bibnamefont{Cristanti}},
  \bibinfo{author}{\bibfnamefont{G.}~\bibnamefont{Paladin}}, \bibnamefont{and}
  \bibinfo{author}{\bibfnamefont{A.}~\bibnamefont{Vulpiani}},
  \emph{\bibinfo{title}{Products of Random Matrices in Statistical Physics}}
  (\bibinfo{publisher}{Springer-Verlag}, \bibinfo{year}{1993}).

\bibitem[{\citenamefont{Dellago and Posch}(1997)}]{posch2}
\bibinfo{author}{\bibfnamefont{C.}~\bibnamefont{Dellago}} \bibnamefont{and}
  \bibinfo{author}{\bibfnamefont{H.~A.} \bibnamefont{Posch}},
  \bibinfo{journal}{Physica~A} \textbf{\bibinfo{volume}{68}},
  \bibinfo{pages}{240} (\bibinfo{year}{1997}).

\bibitem[{\citenamefont{{de Wijn} and {van Beijeren}}({\natexlab{b}})}]{tbp}
\bibinfo{author}{\bibfnamefont{A.~S.} \bibnamefont{{de Wijn}}}
  \bibnamefont{and} \bibinfo{author}{\bibfnamefont{H.}~\bibnamefont{{van
  Beijeren}}}, \bibinfo{note}{to be published.}

\bibitem[{\citenamefont{Searles et~al.}(1997)\citenamefont{Searles, Evans, and
  Isbister}}]{searles}
\bibinfo{author}{\bibfnamefont{D.~J.} \bibnamefont{Searles}},
  \bibinfo{author}{\bibfnamefont{D.~J.} \bibnamefont{Evans}}, \bibnamefont{and}
  \bibinfo{author}{\bibfnamefont{D.~J.} \bibnamefont{Isbister}},
  \bibinfo{journal}{Physica~A} \textbf{\bibinfo{volume}{96}}
  (\bibinfo{year}{1997}).

\end{thebibliography}

\end{document}